\documentclass[onecolumn,showpacs,preprintnumbers,amsmath,amssymb,elsart]{revtex4}

\usepackage{mathrsfs}

\usepackage{graphicx}
\usepackage{dcolumn}
\usepackage{bm}
\usepackage[center]{subfigure}

\begin{document}


\title{Arrayed and checkerboard optical waveguides controlled by the electromagnetically-induced transparency}

\author {Yongyao Li$^{1,2}$}
\email{yongyaoli@gmail.com}
\author{B. A. Malomed$^{3}$}
\author{Mingneng Feng$^{1}$}
\author{Jianying Zhou$^{1}$}
\email{stszjy@mail.sysu.edu.cn}

\affiliation{$^{1}$State Key Laboratory of Optoelectronic Materials
and Technologies,\\Sun Yat-sen University, Guangzhou 510275, China\\
$^{2}$Department of Applied Physics, South China Agricultural
University, Guangzhou 510642, China \\
$^{3}$Department of Interdisciplinary Studies, Faculty of
Engineering, Tel Aviv University, Tel Aviv, Israel}


\begin{abstract}
We introduce two models of quasi-discrete optical systems: an array
of waveguides doped by four-level N-type atoms, and a nonlinear
checkerboard pattern, formed by doping with three-level atoms of the
$\Lambda$-type. The dopant atoms are driven by external fields, to
induce the effect of the electromagnetically-induced transparency
(EIT). These active systems offer advantages and addition degrees of
freedom, in comparison with ordinary passive waveguiding systems. In
the array of active waveguides, the driving field may adjust linear
and nonlinear propagation regimes for a probe signal. The nonlinear
checkerboard system supports the transmission of stable spatial
solitons and their "fuzzy" counterparts, straight or oblique.
\end{abstract}

\maketitle

\section{INTRODUCTION}

The transmission of light in discrete arrays of evanescently coupled
waveguides is a topic of great interest in optics. The arrays are
prime examples of systems in which the discrete optical dynamics can
be observed and investigated \cite{Lederer}. Optical fields
propagating in such settings exhibit a great number of novel
phenomena \cite{Lederer,Kartashov1,Schwartz}. However, traditional
coupled arrays or lattices, such as arrays of waveguides made of
AlGaAs \cite{yaron1} or periodically poled lithium niobate (PPLN)
\cite{Iwanow}, \emph{virtual lattices} created in photorefractive
crystals (PhCs) \cite{Fleischer2} and liquid crystals
\cite{Fratalocchi} by means of the optical induction, \emph{etc.},
are built of passive ingredients.

On the other hand, it is well known that active elements, such as
atoms with a near-resonant transition frequency, may lend the medium
a number of specific optical characteristics, such as strong
dispersion, a complex dielectric constant, and the strong variation
of the dispersion relation near the resonance. Therefore, waveguide
arrays made of active elements may offer low thresholds, in
comparison with their passive counterparts, and possibilities for
the ``management" of their waveguiding characteristics. In the 1D
(one-dimensional) case, active discrete systems were previously
studied in detail in the form of \emph{resonantly-absorbing Bragg
reflectors} (RABRs) \cite{Kurizki}, which are used to demonstrate
optical switching \cite{Prineas}, storage \cite{JYZhou}, and
nonlinear conversion \cite{LJT2006}.

Recently, a 2D (two-dimensional) ``imaginary-part photonic crystal"
(IPPhC, i.e., a medium with a periodic variation of the imaginary
part of the refractive index) is realized by means of the techniques
of multi-beam-interference holography, lithography and back-filling
\cite{LJT1}, which allows one to create a spatially-structured
distribution of the active material. For example, active substance
Rhodamine B can be doped into the homogeneous SU8 background to form
an IPPhC. In this structure, the real part of the refractive index
is constant if the probe wavelength is far detuned from the
resonance. However, the imaginary part of refractive index affects
the real part, which becomes a conspicuous effect close to the
resonance. Thus, in the vicinity the absorption window, the IPPhC
also acts as a traditional PhC. Very recently, a laser system in a
medium featuring a periodic distribution of loss, which is akin to
IPPhC, is demonstrated in the experiment \cite{WXYYYL}.

In this work we propose two new varieties of active light-guiding
systems. In Section II, we show that the introduction of an active
material into the PhC provides for a way to create active structures
in the form of a coupled waveguide arrays. The difference of this
system from the RABR is that guides light not \emph{across} the
periodic structure, but rather \emph{along} it. In Section III, we
introduce a \emph{checkerboard} system, which is built of
alternating linear and the nonlinear square cells in the $x$-$z$
plane. These systems may be controlled (``managed") via the effect
of the electromagnetic-induced transparency (EIT).

\section{Waveguiding arrays controlled by the electromagnetically-induced transparency}
In this section we consider the possibility to use N-type
near-resonant four-level atoms as the active dopant. The respective
scheme of the energy levels is shown in FIG. 1(a), where $|1\rangle$
and $|2\rangle$ are the ground and a metastable states,
respectively. These two states have the same parity of their wave
functions, which is opposite to that of states $|3\rangle$ and
$|4\rangle$.

As a part of the scheme, we assume that a weak probe wave, $E_{P}$,
with Rabi frequency $\Omega_{P}=\wp_{31}E_{P}/\hbar$ is acting on
transition $|1\rangle\rightarrow|3\rangle$, with single-photon
detuning $\Delta_{1}$. Here $\wp_{31}$ (which is assumed real) is
the matrix element of the dipole transition between $|1\rangle$ and
$|3\rangle$ . Further, a traveling-wave field with Rabi frequency
$\Omega_{C}$ drives the atomic transition
$|2\rangle\rightarrow|3\rangle$ with detuning
$\Delta_{C}=\Delta_{1}$, hence the two-photon detuning is given by
$\delta=\Delta_{1}-\Delta_{C}\equiv0$. As another ingredient of the
EIT scheme, an optical-induction field with Rabi frequency
$\Omega_{S}$ induces transition $|2\rangle\rightarrow|4\rangle$,
with detuning $\Delta_{2}$. The decay rate for level $|n\rangle$ is
$\gamma_{n}$. Here we neglect $\gamma_{1}$ and $\gamma_{2}$, and
assume $\gamma_{3}\approx\gamma_{4}\equiv\gamma$.

The Hamiltonian of the system is:
\begin{eqnarray}
H&&=\sum^{4}_{l=1}\hbar\omega_{l}|l\rangle\langle
l|-{1\over2}[\Omega_{P}e^{-i\omega_{P}t}|3\rangle\langle1|+\nonumber\\
&&\Omega_{C}e^{-i\omega_{C}t}|3\rangle\langle2|+\Omega_{S}e^{-i\omega_{S}t}|4\rangle\langle2|+H.C].
\end{eqnarray}
where $\omega_{l}$ is the eigenfrequency of the \emph{l}-th level.
The EIT effect means that, when the probe is exactly at the
two-photon resonance ($\delta=0$), and the atoms are prepared in the
ground state, the linear absorption of the probe vanishes,
irrespective of the single-photon detuning \cite{Fleischhauer2005}.

The quasi-discrete system, which is introduced in this section, is
illustrated by panel (b1) in FIG. 1, which shows the distribution of
the concentration of the active component in the 1D case. The
transverse width of the waveguide is $d_{1}$, the interval between
the waveguides is $d_{2}$, and $N_{0}$ is the density of the N-type
atoms inside the waveguides. Coefficients are chosen as per
experimental in the case of Y$_{2}$SiO$_{5}$ doped with Pr$^{3+}$
(Pr:YSO) \cite{Ham}: the density of active atoms inside the
waveguides is $N_{0}=1.0\times10^{18}$cm$^{3}$ (which corresponds to
the dopant concentration $\approx$0.1\%),
$\wp_{31}=1.18\times10^{-32}$C$\cdot$m, and $\gamma=30$ kHz, the
probe wavelength being 605 nm. Linear and the nonlinear properties
of this system are detailed below.

\textbf{(1) Linear properties of the system}

If we set $\Delta_{1}=0$ and assume $\Delta_{2}\gg\Omega_{C},\gamma$
, then the absorption of the probe can be neglected. One can easily
find the one-step steady-state solution for the density-matrix
element of the transition between $|1\rangle$ and $|3\rangle$ , cf.
Ref. \cite{Lukin}:
\begin{eqnarray}
\rho_{31}={|\Omega_{S}|^{2}\over2\Delta_{2}|\Omega_{C}|^{2}}\Omega_{P}.
\end{eqnarray}

\begin{figure} 
\centering \subfigure[]{
\label{fig_1_a} 
\includegraphics[scale=0.18]{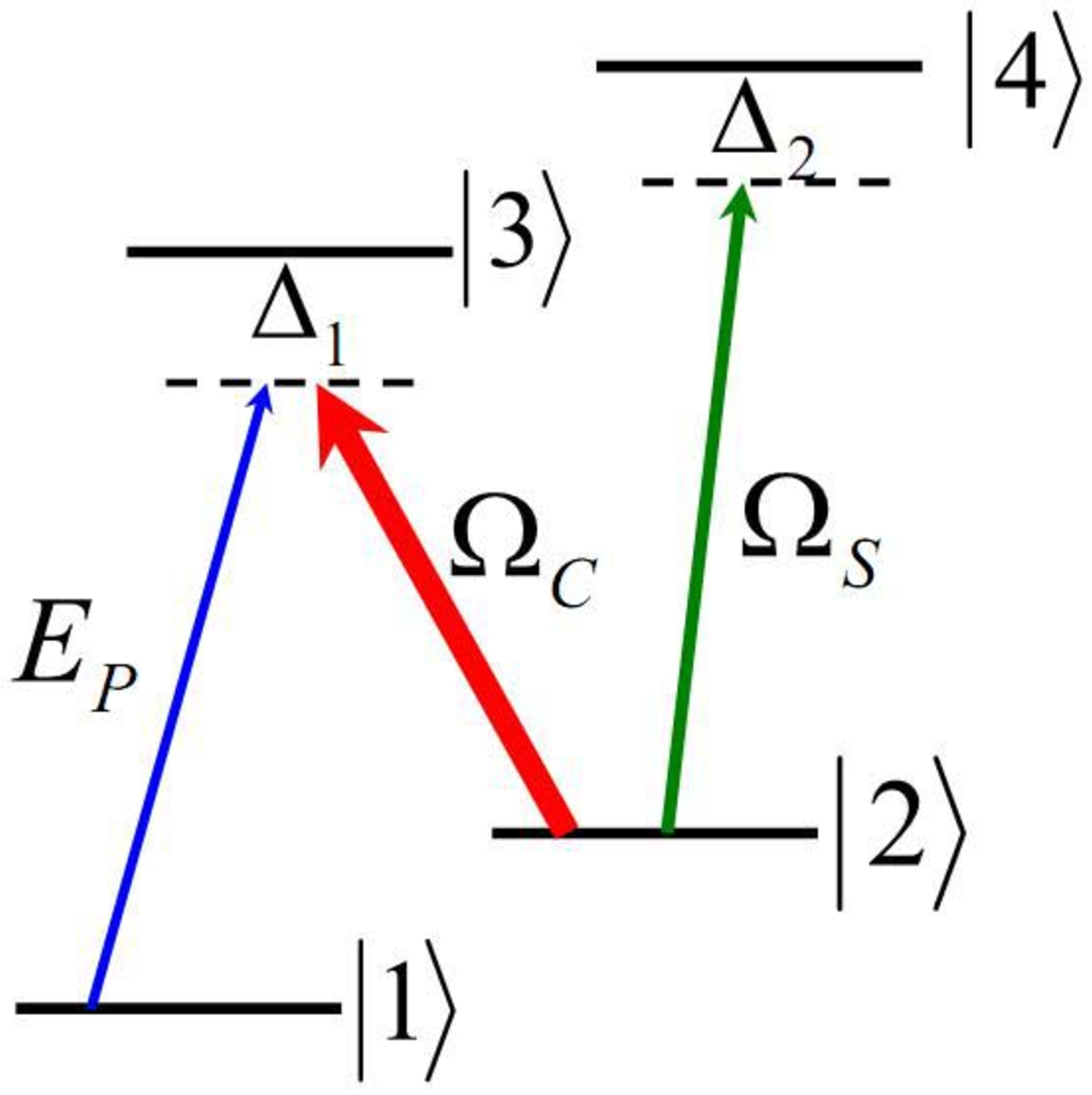}}
\hspace{0.02in} \subfigure[]{
\label{fig_2_b} 
\includegraphics[scale=0.12]{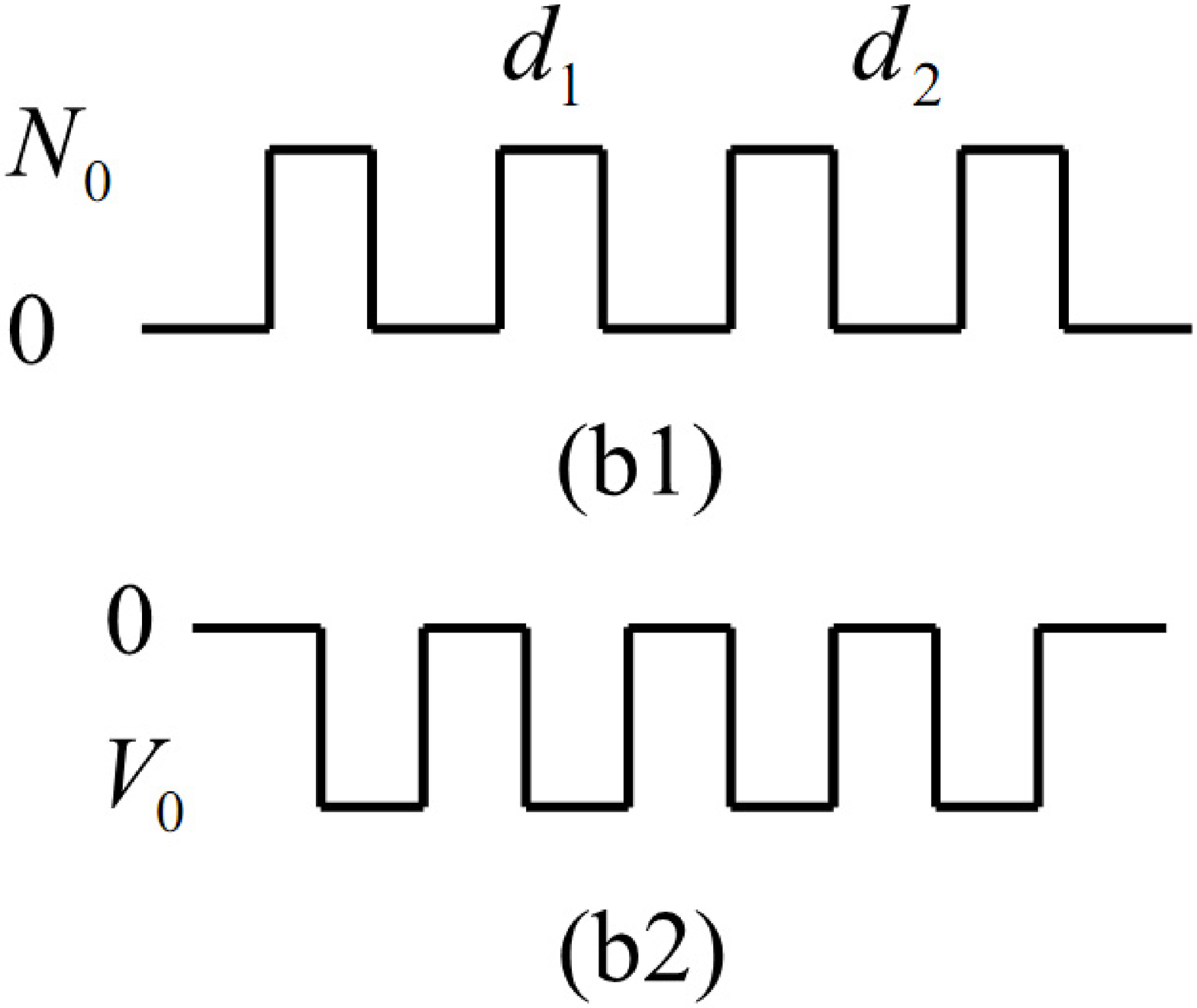}}
\hspace{0.02in} \subfigure[]{
\label{fig_2_b} 
\includegraphics[scale=0.15]{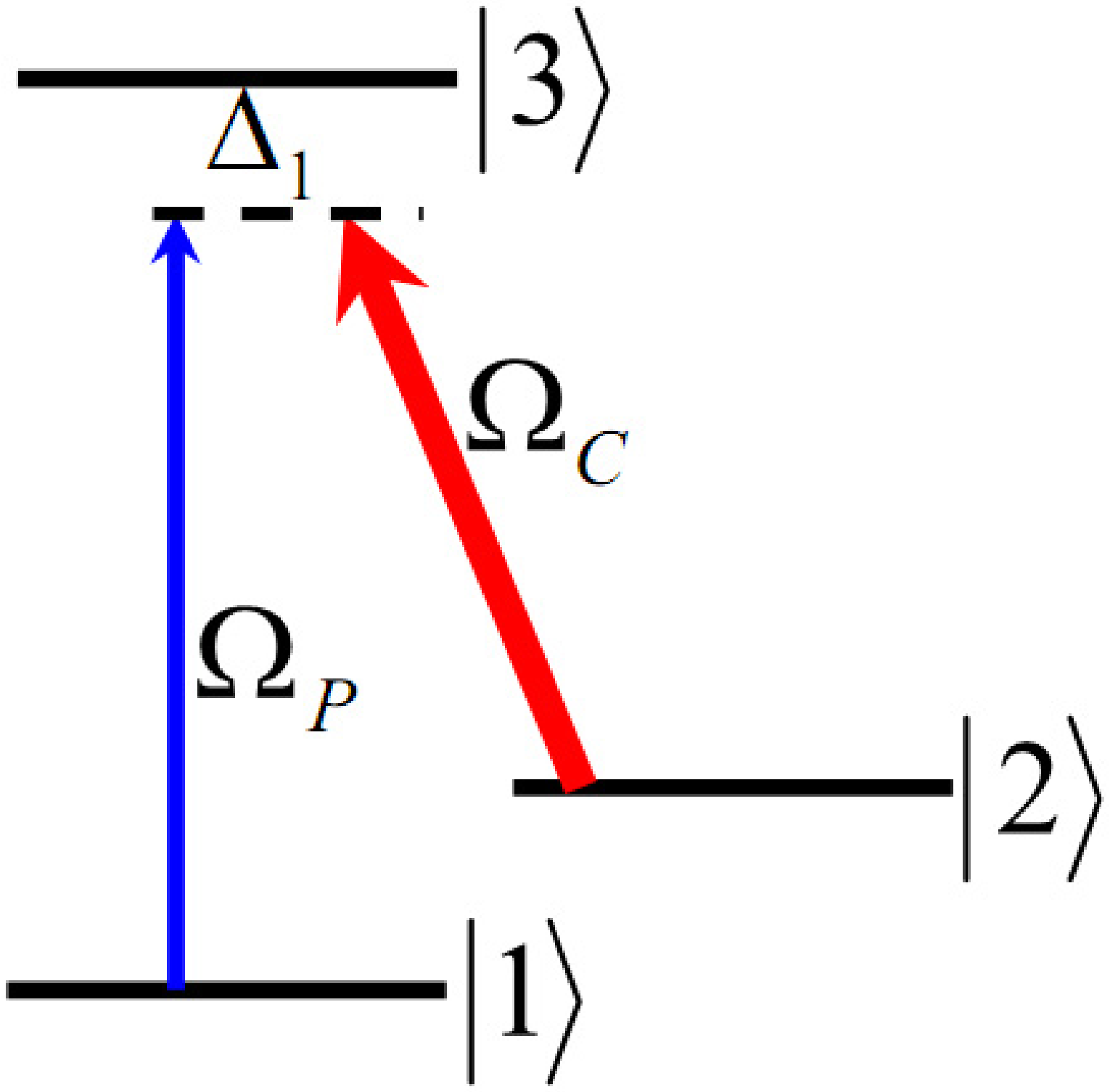}}
\hspace{0.02in} \subfigure[]{
\label{fig_2_b} 
\includegraphics[scale=0.12]{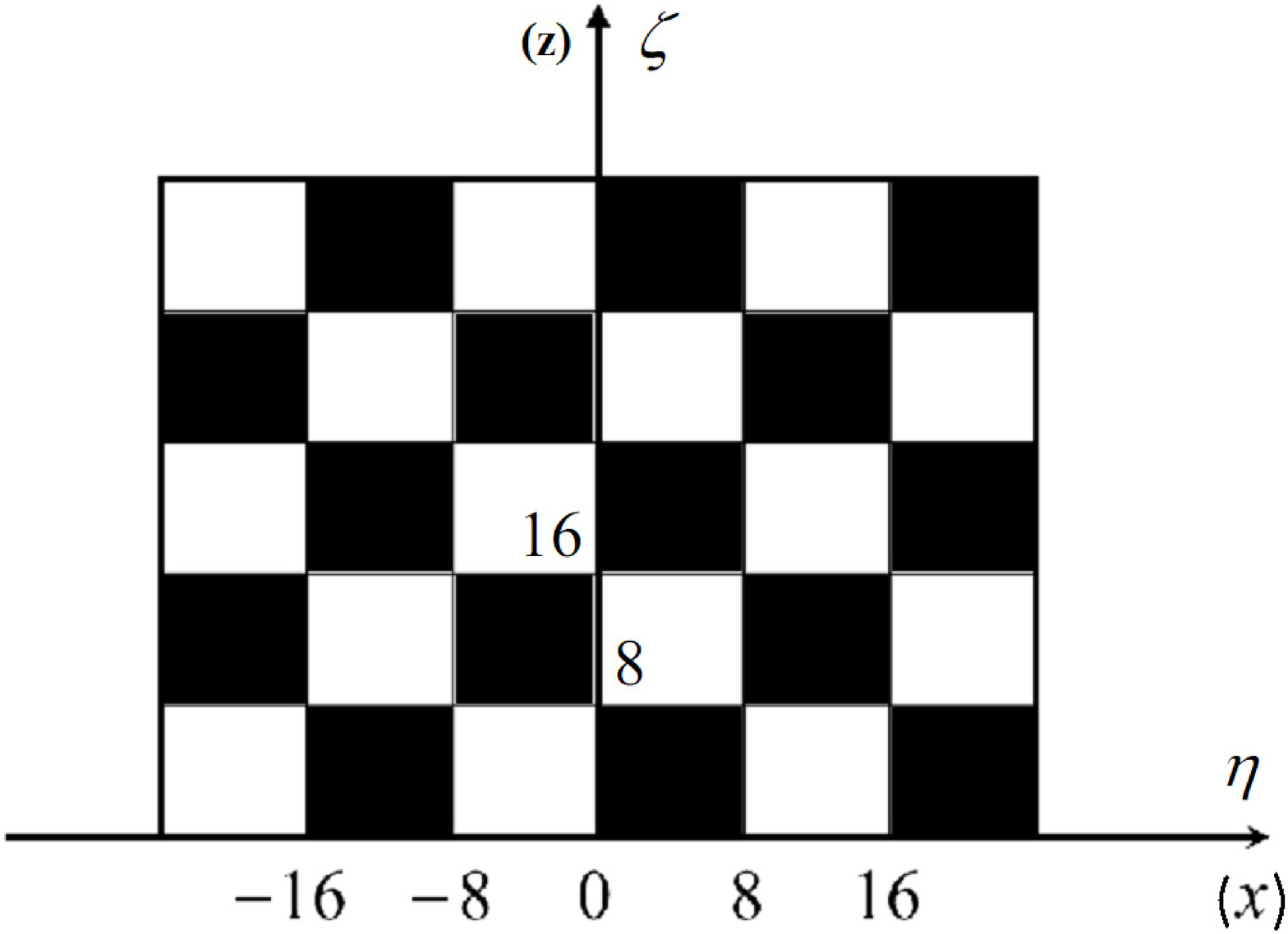}}
\caption{\label{fig:wide} (Color online) (a) The energy-level
diagram of the N-type atom. (b1) The structure of the active
waveguide array: the transverse width of the waveguide is $d_{1}$ ,
the interval between the waveguides is $d_{2}$, and the density of
the active atoms in the waveguide is $N_{0}$ . (b2) The respective
effective periodic potential with
$V_{0}=-N_{0}\wp^{2}_{31}|\Omega_{S}|^{2}/2\epsilon_{0}\hbar\Delta_{2}|\Omega_{C}|^{2}$
. (c) The energy-level diagram of the $\Lambda$-type atom. (d) The
checkerboard system: black and white square cells depict areas which
are, respectively, doped with the active material, or are left
undoped.}\label{fig_1}
\end{figure}

Therefore, the contribution of the resonant atoms to the
polarization experienced by the probe is
$\mathscr{P}=2N\wp_{31}\rho_{31}$, cf. Ref. \cite{Scully} (recall
$N$ is the dopant density). The (1+1)D paraxial propagation equation
for the slowly varying envelope of the probe field, $E_{P}$, is
\begin{eqnarray}
2ik_{P}{\partial\over\partial z}E_{P}=-{\partial^{2}\over\partial
x^{2}}E_{P}-{k^{2}_{P}\over\epsilon_{0}}\mathscr{P},
\end{eqnarray}
where $k_{P}=2\pi n/\lambda_{P}$ is the wavenumber of the probe, and
$n$ the background refractive index. The substitution of Eq. (2)
into the last term of Eq. (3) yields a scaled linear Schr\"{o}dinger
equation,
\begin{eqnarray}
i{\partial\over\partial\zeta}U=-{1\over2}{\partial^{2}\over\partial\eta^{2}}U+V(\eta)U,
\end{eqnarray}
where $\zeta=k_{p}z$, $\eta=k_{p}x$, $U=\Omega_{P}/\gamma$, and the
effective potential,
\begin{eqnarray}
V(\eta)=-{\wp^{2}_{31}\over2\epsilon_{0}\hbar\Delta_{2}}{|\Omega_{S}|^{2}\over|\Omega_{C}|^{2}}N(\eta).
\end{eqnarray}
is induced by the concentration distribution $N(\eta)$ , as shown in
panel (b1) of FIG. 1. This potential is induced by the giant Kerr
effect controlled by the Rabi frequency $\Omega_{S}$ \cite{Schmidt}.
In fact, this potential emulates a difference in the local
refractive index contrast between the active (doped) and passive
(undoped) regions. For $\Delta_{2}>0$ (here, we choose
$\Delta_{2}=100\gamma$), the shape the potential is shown in panel
(b2) of FIG. 1, which is similar to the Kronig-Penney potentials
corresponding to the tight-binding model in solid-state physics
\cite{Kittel}. The depth of local wells in the periodic potential is
defined by the intensity ratio, $|\Omega_{S}|^{2}/|\Omega_{C}|^{2}$,
the corresponding refractive-index contrast between the active and
passive regions being $\Delta
n=\wp^{2}_{31}N_{0}|\Omega_{S}|^{2}/2\epsilon_{0}\hbar\Delta_{2}|\Omega_{C}|^{2}$

\begin{figure} 
\centering \subfigure[]{
\label{fig_1_a} 
\includegraphics[scale=0.3]{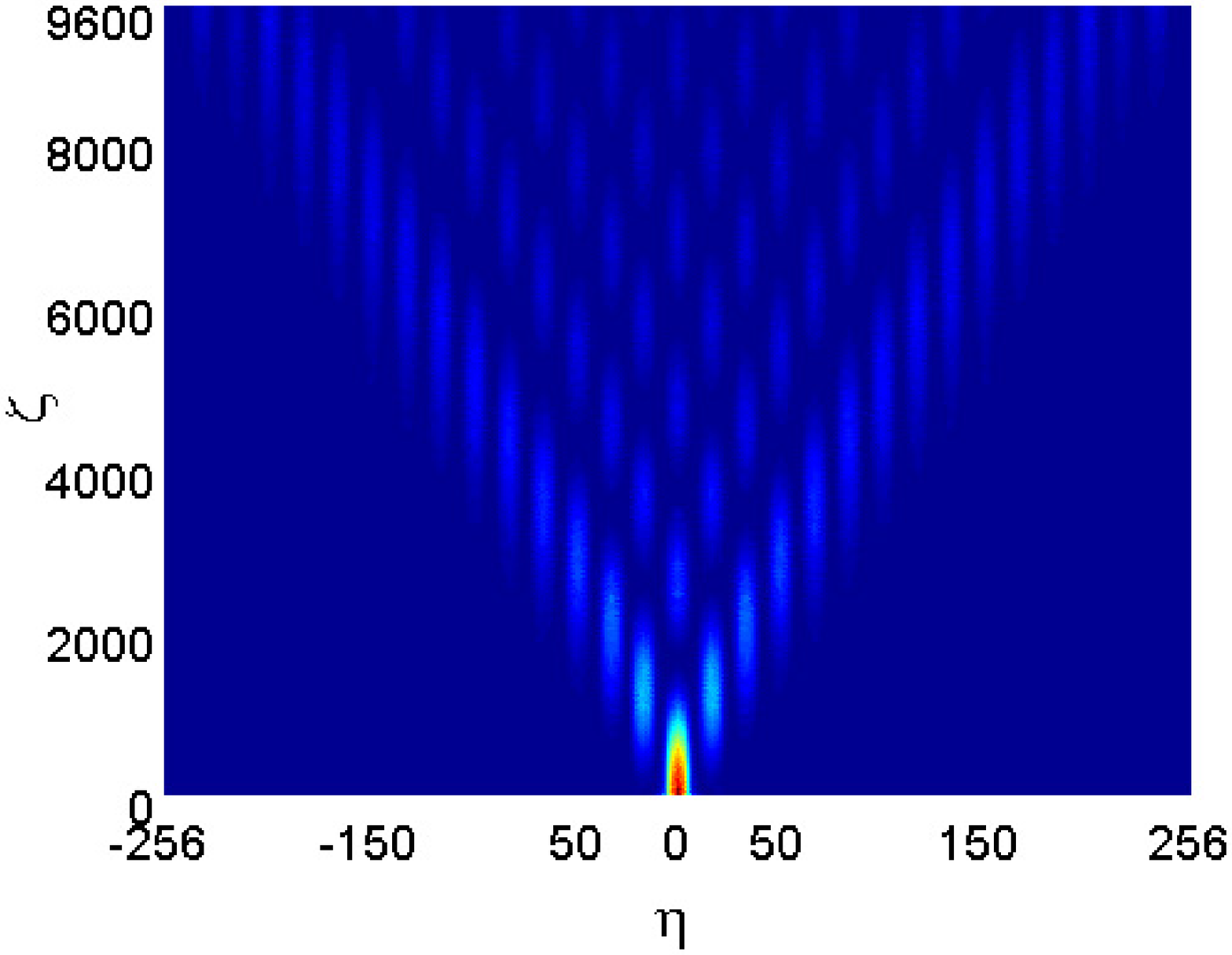}}
\hspace{0.02in} \subfigure[]{
\label{fig_2_b} 
\includegraphics[scale=0.3]{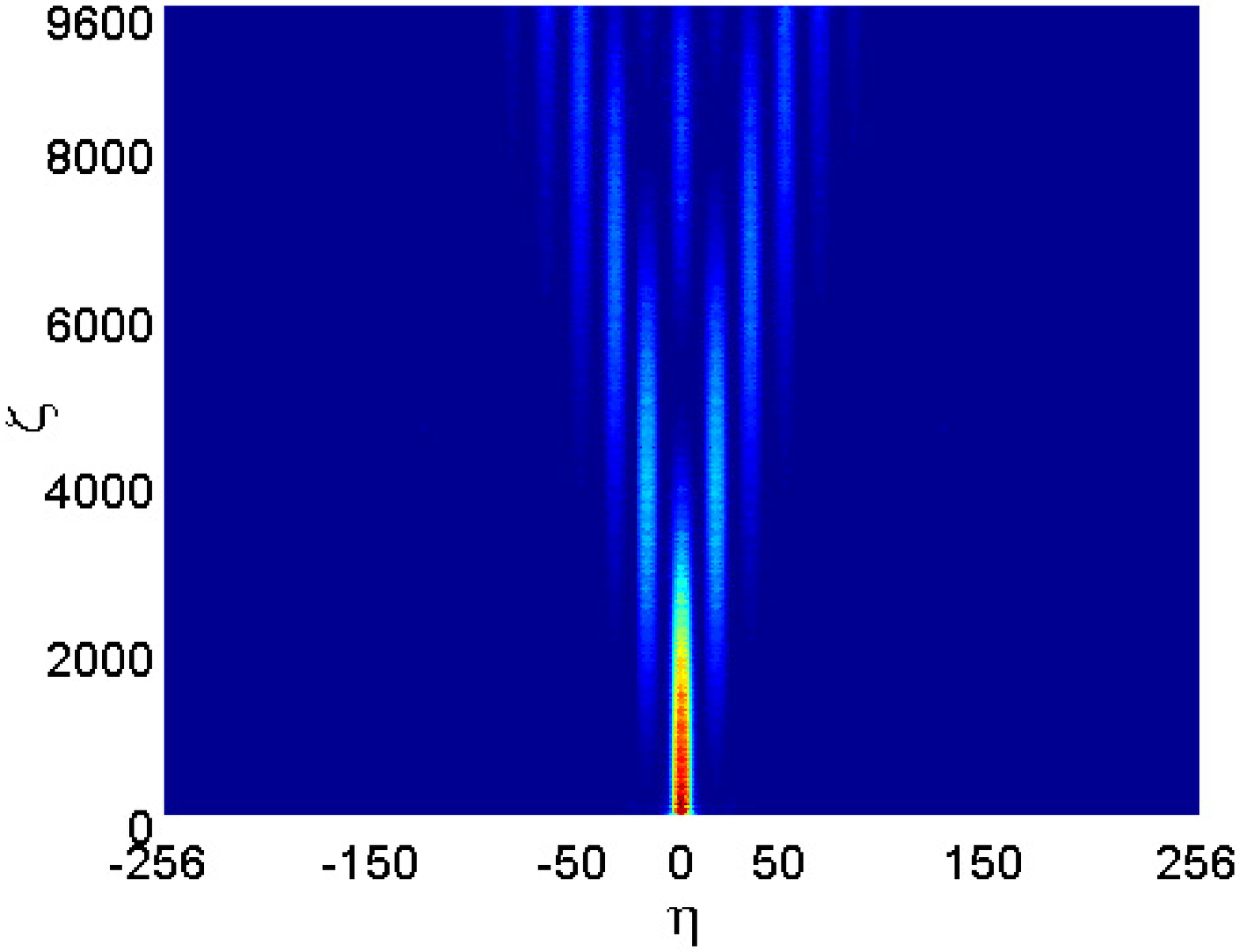}}
\hspace{0.02in} \subfigure[]{
\label{fig_2_b} 
\includegraphics[scale=0.3]{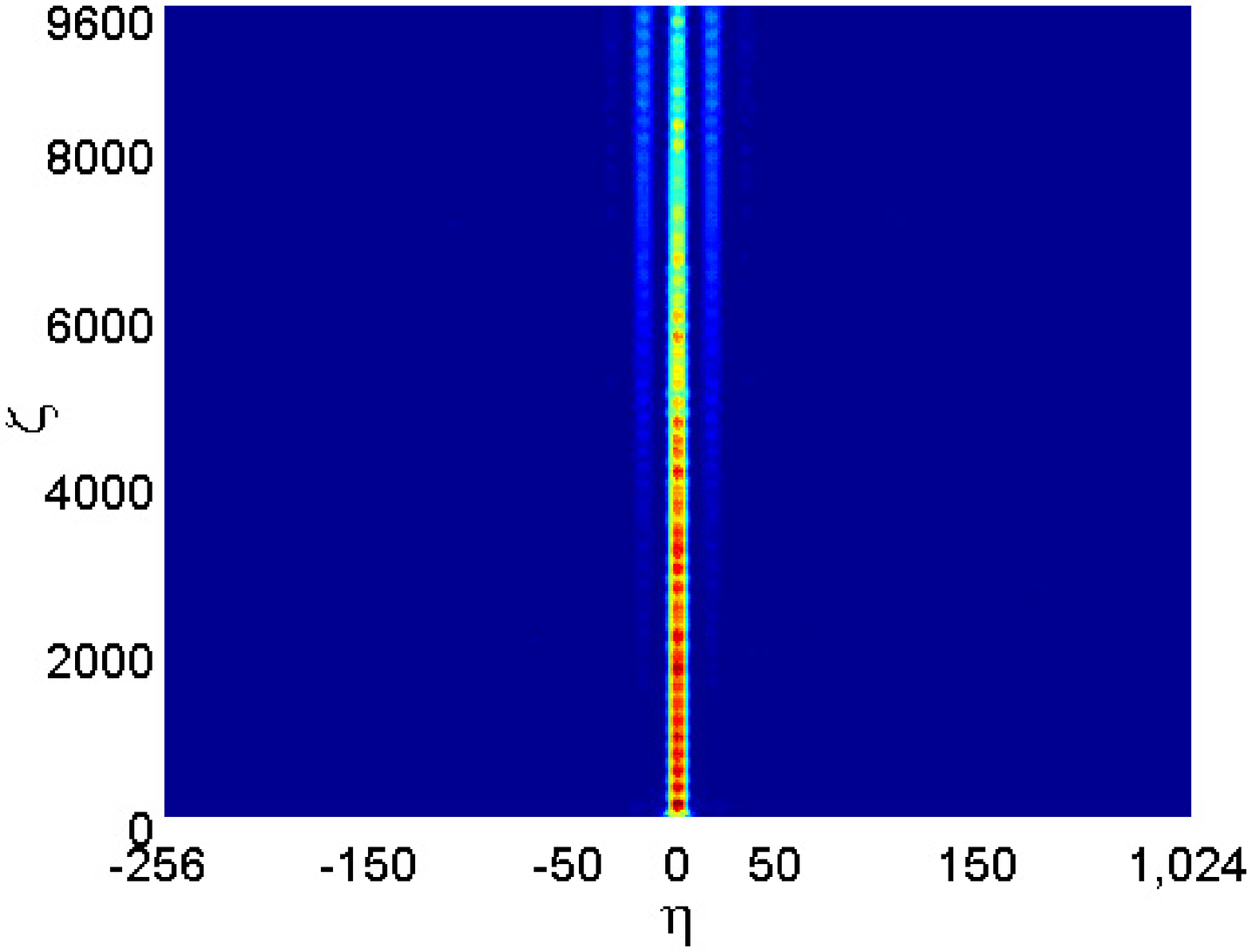}}
\hspace{0.02in} \subfigure[]{
\label{fig_2_b} 
\includegraphics[scale=0.3]{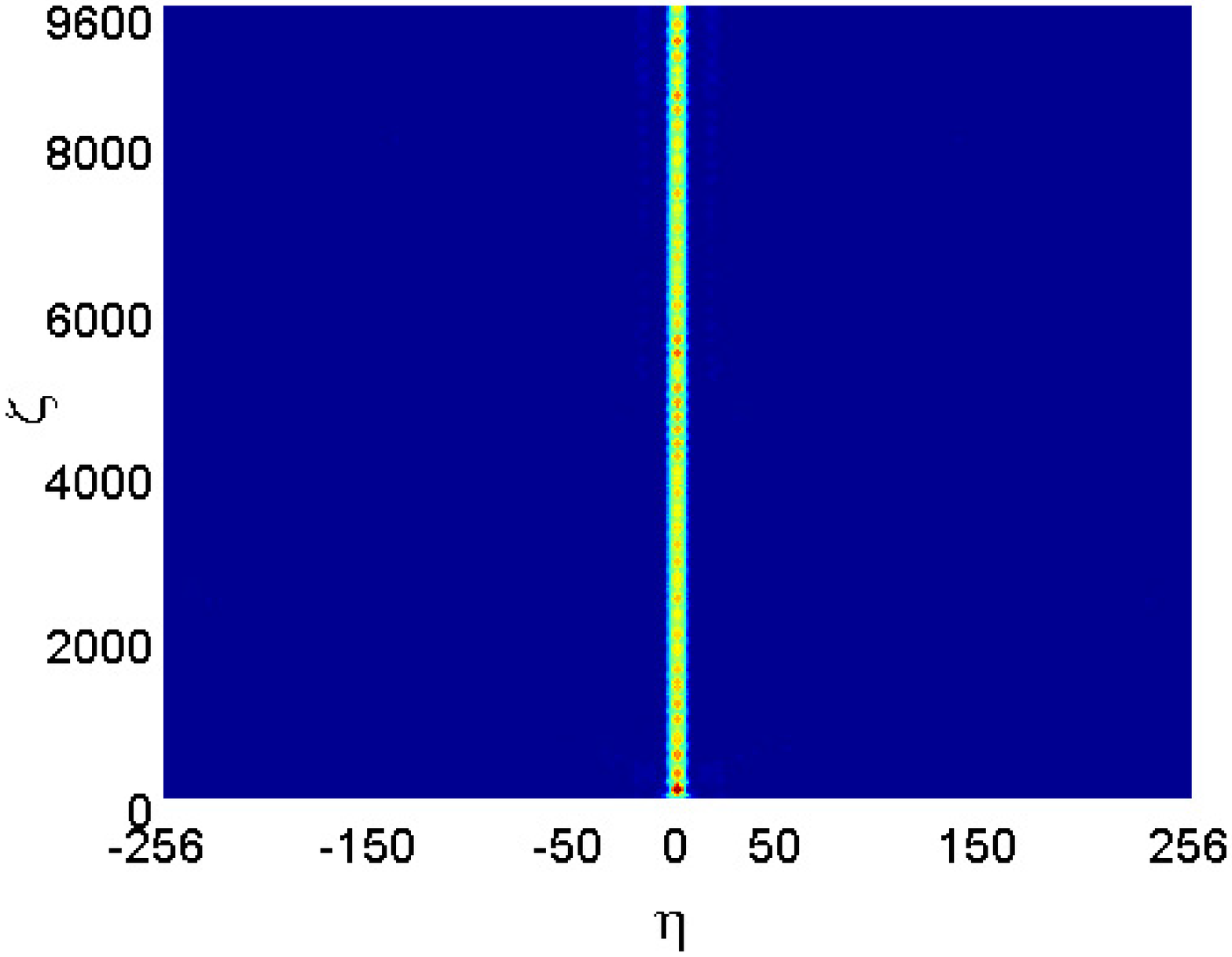}}
\caption{\label{fig:wide}(Color online) Simulations of the linear
transmission of light in the arrays of active waveguides. Here, we
chose $d_{1}=d_{2}=8$, $\Omega_{C}=2\gamma$ and
$\Delta_{2}=100\gamma$. The incident beam (at $\zeta=0$) is
$U=A_{0}\exp(-\eta^{2}/d^{2}_{1})$ with $A_{0}=0.07$. (a)
$|\Omega_{S}|^{2}=0.036|\Omega_{C}|^{2}$ (i.e., $\Delta n=0.09$).
(b) $|\Omega_{S}|^{2}=0.064|\Omega_{C}|^{2}$ (i.e., $\Delta
n=0.16$). (c) $|\Omega_{S}|^{2}=0.100|\Omega_{C}|^{2}$ (i.e.,
$\Delta n=0.25$). (d) $|\Omega_{S}|^{2}=0.144|\Omega_{C}|^{2}$
(i.e., $\Delta n=0.36$). } \label{fig_2}
\end{figure}
Figure 2 displays the propagation of the probe in such a system at
different values of $|\Omega_{S}|^{2}/|\Omega_{C}|^{2}$ (i.e.
different values of $\Delta n$), produced by numerical simulations
of Eq. (4). This figure demonstrates that the quasi-discrete
diffraction naturally gets suppressed when the depth of the
potential, i.e., $|\Omega_{S}|^{2}/|\Omega_{C}|^{2}$ or $\Delta n$,
increases, resulting in a reduced coupling between local waveguides.
Thus, the diffraction in the present setting may be efficiently
controlled by varying the magnitude of
$|\Omega_{S}|^{2}/|\Omega_{C}|^{2}$.

\textbf{(2) Nonlinear properties of the system}

If both detunings satisfy conditions
$\Delta_{1},\Delta_{2}\gg\Omega_{C},\gamma$, then the absorption of
the probe (the losses) may be neglected. Moreover $\Delta_{1}\neq0$
produces an enhanced Kerr nonlinearity \cite{XiaoM}, provided that
the density matrix element $\rho_{31}$, which accounts for
transitions between $|1\rangle$ and $|3\rangle$, is taken into
regard, to describe the third-order effect. The steady-state
solution for this matrix element can be obtained in the following
from, cf. Ref. \cite{liyongyao1}:

\begin{eqnarray}
\rho_{31}=\rho^{(1)}_{31}+\rho^{(2)}_{31}+\rho^{(3)}_{31}\approx{|\Omega_{S}|^{2}\over2\Delta_{2}|\Omega_{C}|^{2}}\Omega_{P}-{|\Omega_{P}|^{2}\over2\Delta_{1}|\Omega_{C}|^{2}}\Omega_{P}
\end{eqnarray}
With regard to this result and the above definition,
$U=\Omega_{P}/\gamma$, Eq. (3) changes its form into that of the
standard nonlinear Schr\"{o}dinger (NLS) equation \cite{Kevrekidis}:
\begin{eqnarray}
i{\partial\over\partial\zeta}U=-{1\over2}{\partial^{2}\over\partial\eta^{2}}U+V(\eta)U+\kappa(\eta)|U|^{2}U,
\end{eqnarray}
where the effective nonlinear coefficient is
\begin{eqnarray}
\kappa(\eta)={\wp^{2}_{31}\over2\epsilon_{0}\hbar\Delta_{1}(|\Omega_{C}|^{2}/\gamma^{2})}N(\eta)
\end{eqnarray}

Thus, $E_{P}$ is affected by the modifications of the refractive
index of two types: (1) a periodic change of the linear index
induced by $\Omega_{S}$ via the giant Kerr effect; (2) the nonlinear
change under the action of $E_{P}$ itself, via the enhanced
self-Kerr effect. The sign of detuning $\Delta_{1}$ determines
whether the latter effect gives rise to the self-focusing
($\Delta_{1}<0$) or self-defocusing ($\Delta_{1}>0$) sign of the
nonlinearity. When the tunneling coupling between adjacent
waveguides is balanced by the nonlinearity, quasi-discrete solitons
\cite{Mayteevarunyoo} can be formed in this system, which is similar
to those in the traditional coupled arrays made of passive materials
\cite{Kominis}.

\section{Solitons in the checkerboard system controlled by the electromagnetically-induced transparency}

In this section, we assume that the active material is doped
periodically in both the transverse and propagation directions
(i.e., along the \emph{x} and \emph{z} axes, respectively). The
corresponding density distribution of the active material is
$N(x,z)=N_{0}R(x,z)$, where $R(x,z)$ is a dimensionless structural
function of the distribution. Here, we adopt for $R(x,z)$ the form
checkerboard form depicted in FIG. 1(d). The white cells, with
$R(x,z)=0$ , are areas which are not subject to the doping, while
black cells, with $R(x,z)=1$ , depict areas doped by the active
material.

Note that the formation of 2D spatial solitons in the
checkerboard-shaped linear potential was considered in Ref.
\cite{Malomed}, assuming that the probe beam was shone along the
uniform direction in the bulk medium equipped with the transverse
checkerboard structure. Here, the difference is that the modulation
is applied to the \emph{nonlinear} term, and light propagated
\emph{across} the structure.

If we turn off the control field $\Omega_{S}$, the four-level N-type
atomic configuration reduces to the three-level one of the
$\Lambda$-type, see FIG. 1(c). In this case, the
linear-refractive-index contrast between the active and the passive
areas vanishes, which leaves only the nonlinearity modulation in
action.

Again assuming that the detuning is much larger than the decay rate,
$\Delta_{1}\gg\gamma$ , the absorption of the probe may be ignored
as before. Accordingly, in the present setting Eq. (6) is rewritten
as:
\begin{eqnarray}
\rho_{31}=\rho^{(1)}_{31}+\rho^{(2)}_{31}+\rho^{(3)}_{31}\approx-{|\Omega_{P}|^{2}\over2\Delta_{1}|\Omega_{C}|^{2}}\Omega_{P}
\end{eqnarray}
and Eq. (3) changes into
\begin{eqnarray}
i{\partial\over\partial\zeta}U=-{1\over2}{\partial^{2}\over\partial\eta^{2}}U+\kappa(\eta,\zeta)|U|^{2}U,
\end{eqnarray}
where we define
\begin{eqnarray}
\kappa(\eta,\zeta)={\wp^{2}_{31}N_{0}\over2\epsilon_{0}\hbar\Delta_{1}(|\Omega_{C}|^{2}/\gamma^{2})}R(\eta,\zeta)\equiv\kappa_{0}R(\eta,\zeta)
\end{eqnarray}
Therefore, the white and black cells, with $\kappa(\eta,\zeta)=0$,
and $\kappa(\eta,\zeta)=\kappa_{0}$, act as linear and nonlinear
elements, respectively.

The periodic modulation of the nonlinearity in Eq. (10) places this
equation into the broad class of models with nonlinear lattices (see
original works \cite{Sakaguchi} and review \cite{Kartashov}).
However, the present checkerboard pattern of the modulation in the
longitudinal and transverse direction was not studied in previous
works. More general patterns, that should be studied separately, may
be represented by arrays of isolated black squares set against the
white background, or vice versa.

Below, we choose values $\Delta_{1}=-100\gamma$,
$\Omega_{C}=0.5\gamma$, with the other parameters taken as before.
This yields the nonlinearity-modulation amplitude $\kappa_{0}=-10$ .
The probe field is launched at $\zeta=0$ as a Gaussian：
\begin{eqnarray}
U(\eta,0)=A\exp[-(\eta-\alpha)^{2}/W^{2}],
\end{eqnarray}
with the central point at $\eta=\alpha$. For the simulations, we
take the checkerboard with squares of size $8\times8$, as shown in
FIG. 1(d).

The simulations of the evolution of the Gaussian in the framework of
Eq. (10) were carried out by means of the split-step Fourier method.
First, we chose the amplitude and width of Gaussian (12) as,
$A$=0.075, 0.065, 0.055 and $W$=8, with the center placed at the
mid-point of the nonlinear cell ($\alpha=4$). Results of the
simulations are displayed in FIG. 3(a)-(c). In particular, FIG. 3(a)
shows that the probe field with $A$=0.075 propagates without decay
and distortion over the distance longer than $z=100\times W^{2}$,
i.e., $\sim$100 diffraction lengths. This dynamical regime may be
naturally identified as a stable soliton. On the other hand, in FIG.
3 (b), with the Gaussian's amplitude $A$=0.065, the probe field
forms a \emph{fuzzy beam}, which, nevertheless, avoids decay over
the distance exceeding $z=100\times W^{2}$.

However, for $A$=0.055, FIG. 3(c) demonstrates that the input cannot
form a robust beam and rapidly decays. Therefore, there must be
internal borders (i.e. thresholds) separating the stable solitons,
fuzzy beams, and decaying ones, in the plane of the width and
amplitude of the Gaussian inputs. These borders are plotted in FIG.
3(d).

\begin{figure} 
\centering \subfigure[]{
\label{fig_3_a} 
\includegraphics[scale=0.3]{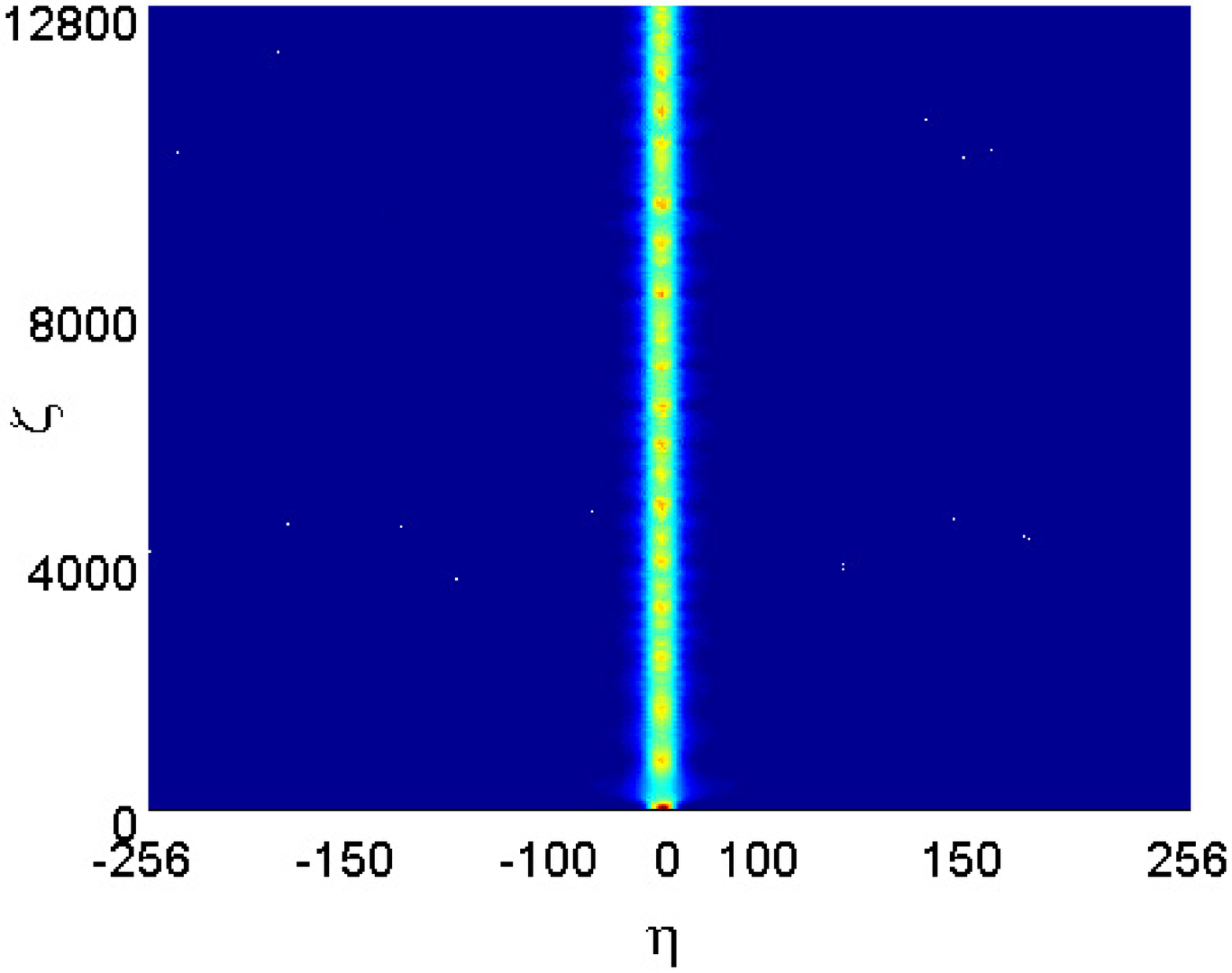}}
\hspace{0.02in} \subfigure[]{
\label{fig_3_b} 
\includegraphics[scale=0.3]{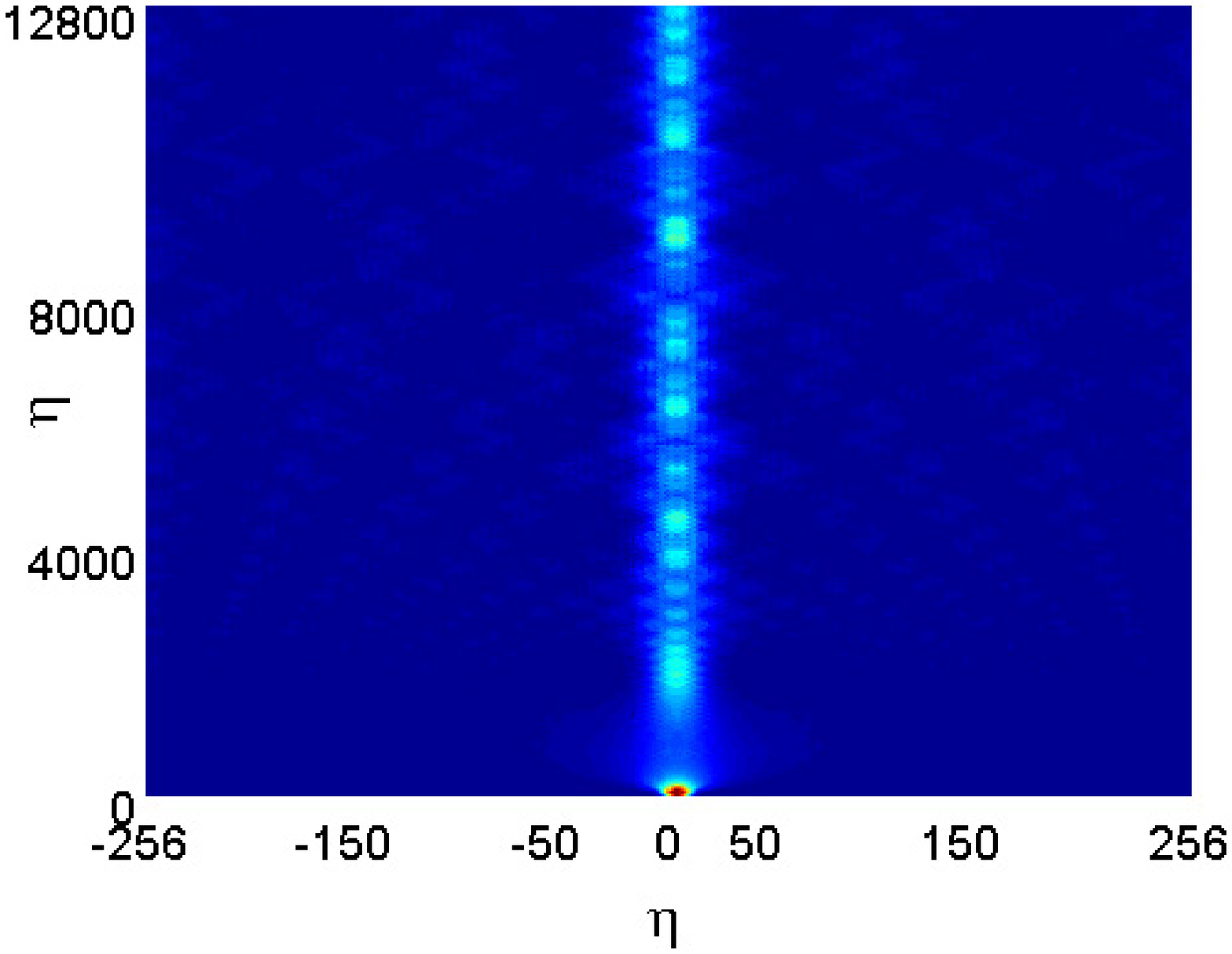}}
\hspace{0.02in} \subfigure[]{
\label{fig_3_c} 
\includegraphics[scale=0.3]{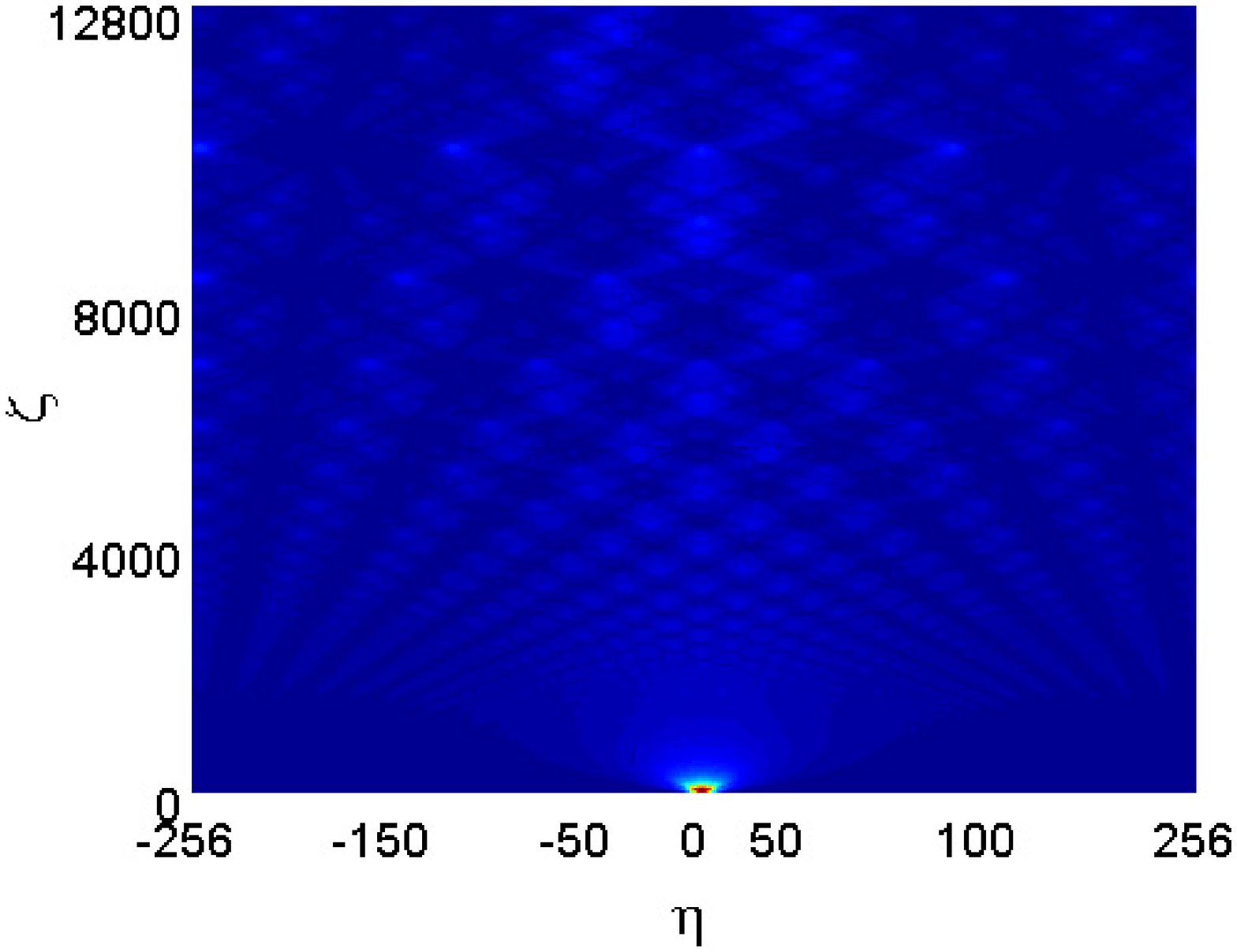}}
\hspace{0.02in} \subfigure[]{
\label{fig_3_d} 
\includegraphics[scale=0.22]{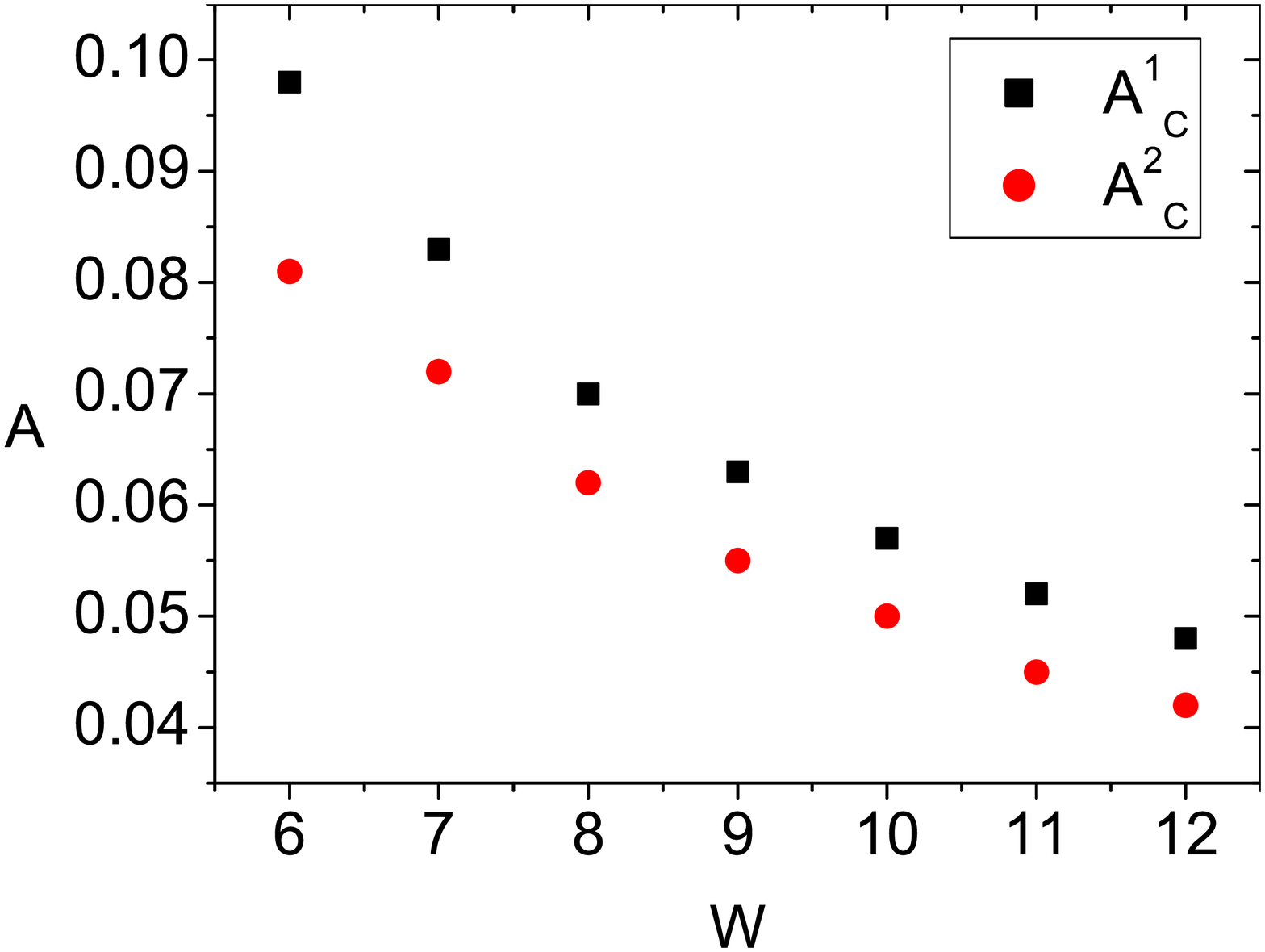}}
\caption{\label{fig:wide}(Color online) Simulations of the evolution
of the Gaussian input probe with width $W=8$ in the checkerboard
model. (a) With amplitude $A$=0.075, the probe propagates in a
stable fashion over the distance of $L=200\times W^{2}=12800$. This
beam may be called a stable soliton. (b) With $A=0.065$, the probe
propagates, keeping a fuzzy shape, still featuring robustness
against the diffractive spreading-out. (c) With $A=0.055$, the probe
quickly spreads out. (d) In the plane of the width and amplitude if
the input Gaussian, $A^{1}_{C}(W)$ is the border between the stable
and fuzzy beams, while $A^{2}_{C}(W)$ is the border between the
fuzzy but robust beams and decaying inputs.
 } \label{fig_3}
\end{figure}
Further simulations, displayed in FIG. 4(a), show that stable
straight solitons can be formed as well if the Gaussian is launched
at the midpoint of the linear cell. However, if the center of the
Gaussian does not coincide with the center of the linear or
nonlinear cell, the propagation of the soliton beam becomes oblique,
see an example in FIG. 4(b).

It may be interesting to consider the oblique propagation of beams
across the checkerboard, induced by the application of a lateral
kick to the input. Another issue of obvious interest is the
interaction of beams in this setting. These generalizations will be
reported elsewhere.

\begin{figure} 
\centering \subfigure[]{
\label{fig_1_a} 
\includegraphics[scale=0.3]{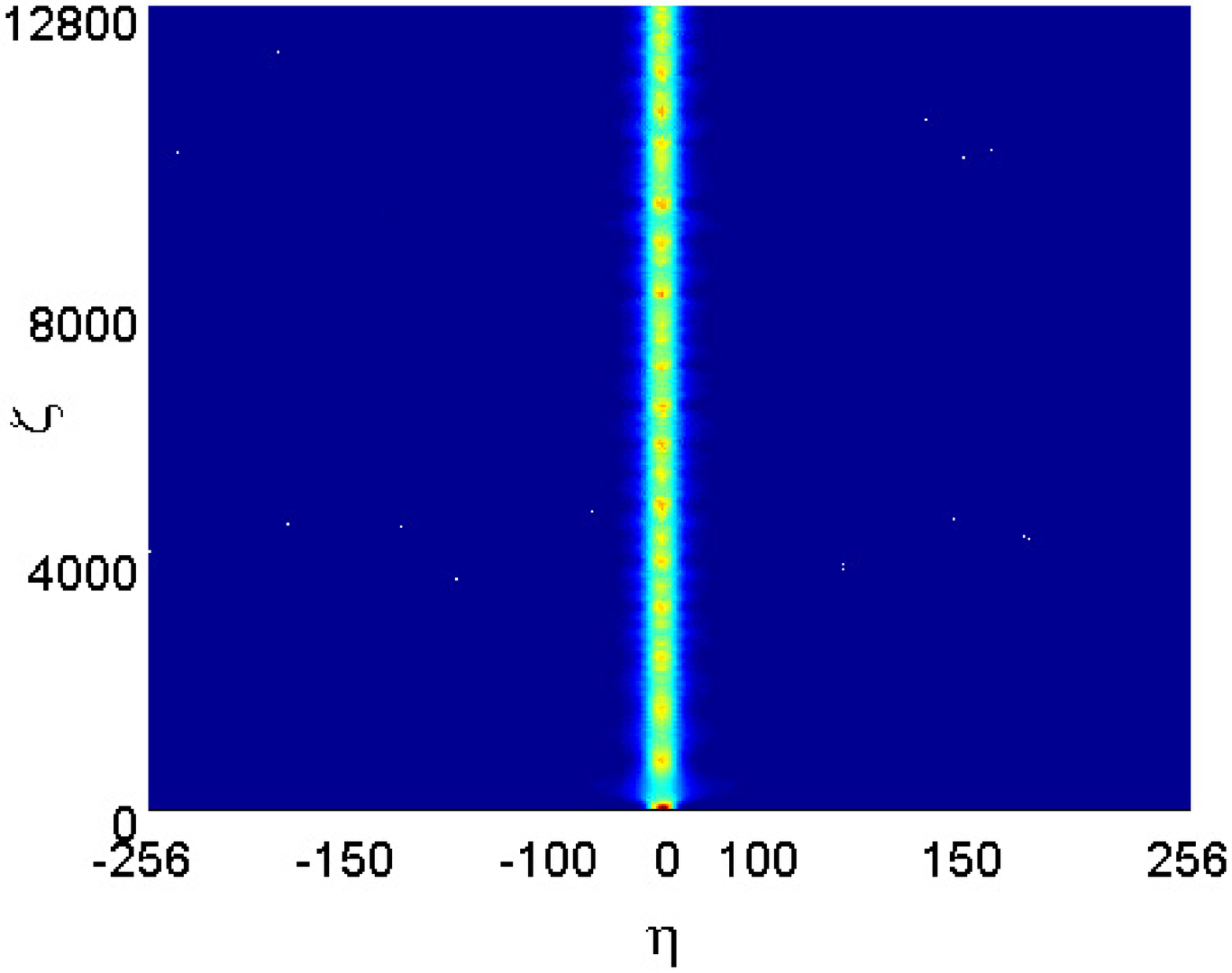}}
\hspace{0.02in} \subfigure[]{
\label{fig_2_b} 
\includegraphics[scale=0.3]{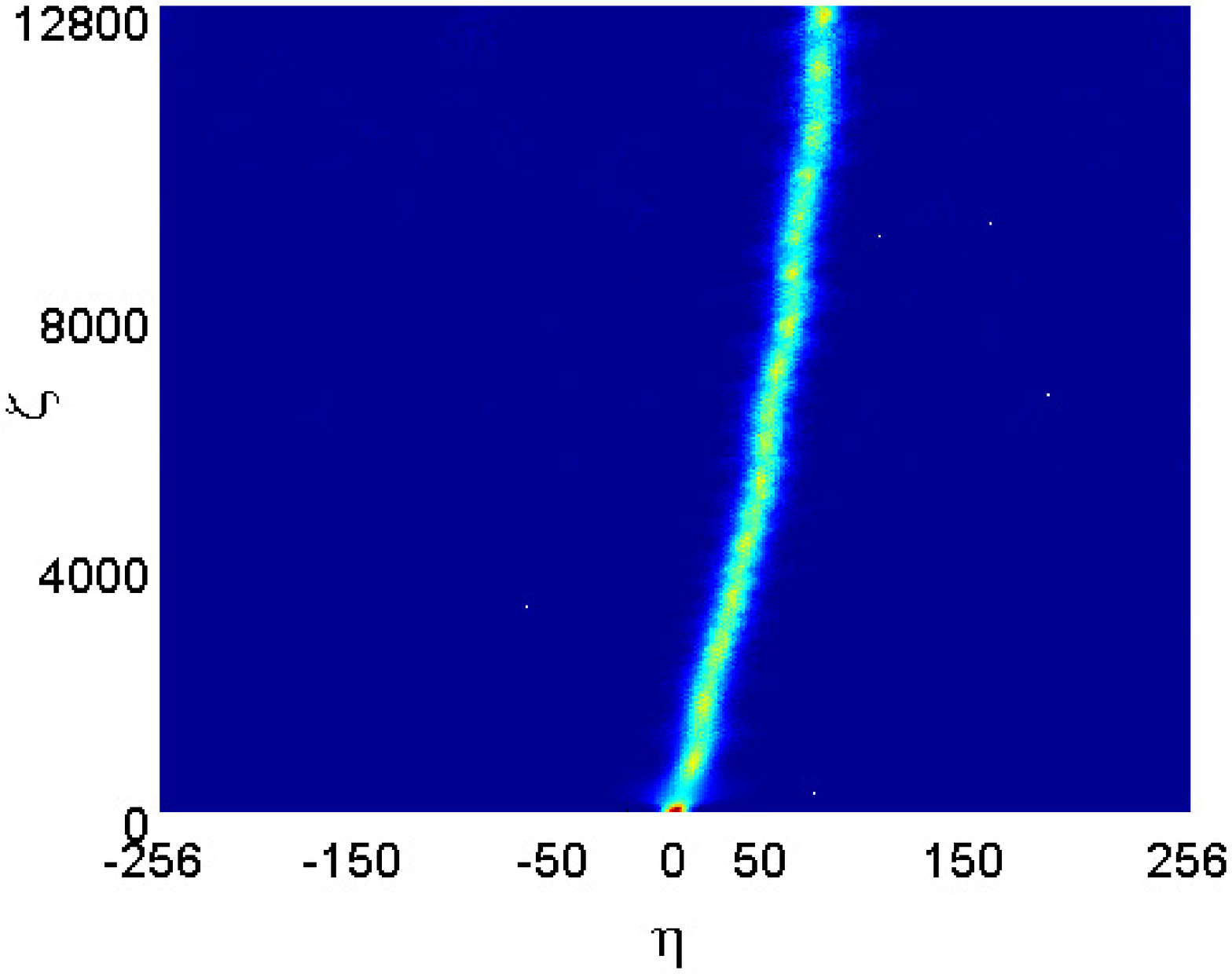}}
\caption{\label{fig:wide} (Color online) (a) The probe with
parameters $A=0.075$, $W=8$, and $\alpha=-4$ is launched at the
mid-point of the linear cell. The probe propagates in a stable
fashion over distance $L=200\times W^{2}=12800$, therefore it is
categorized as a stable soliton. (b) The probe with $A=0.075$,
$W=8$, $\alpha=0$ propagate obliquely if it is launched off the
mid-point of a linear or nonlinear cell. }\label{fig_4}
\end{figure}

\section{CONCLUSION}

In this work, we have proposed a method for building systems of
coupled active waveguides, and also a new system with the
checkerboard pattern of the modulation of the nonlinearity
coefficient. These settings can be created using the appropriate
doping patterns of the N-type four-level and $\Lambda$-type
three-level resonant atoms, respectively, and driving them by means
of the EIT mechanism. Such active systems offer certain advantages,
admitting more possibilities for the design and management, in
comparison to the passive media. Firstly, the nonlinearity in the
active systems can be switched via the sign of detuning
$\Delta_{1}$: the incident beam with $\Delta_{1}<0$ and
$\Delta_{1}>0$  will experience the action of the self-focusing and
self-defocusing nonlinearity, respectively. Next, it is well known
that the EIT may be efficiently applied to few-photon settings,
especially in the nonlinear regimen \cite{Fleischhauer2005,THong}.
Therefore, the systems introduced here may, in principle, offer an
advantage for handling quantum and non-classical light beams,
composed of few photons. Further, it is well known that the group
velocity of the probe can be coherent controlled \cite{lyy3} and
tuned to a very small value under the action of the EIT \cite{Hau},
which implies that the probe signal in this system can be trapped in
the from of the slow light. Thus, various applications of the EIT,
such as dark-state polaritons \cite{lukin2}, the few-photon
four-wave mixing \cite{Fleischhauer2002}, ultra-weak and ultraslow
light \cite{liyongyao2}, \emph{etc.}, may be realized in the systems
proposed here. Furthermore, using properly designed holographic
patterns, various complex spatial structures of the distribution of
the dopant concentration can be photoinduced in the 2D geometry,
such as quasi-crystals \cite{Freedman}, honeycomb lattices
\cite{Peleg}, defect lattices \cite{liyongyao3}, ring lattices
\cite{WXS}, \emph{etc.}, in addition to the simplest checkerboard
patterns analyzed herein. Such 2D structures may have their own
spectrum of potential applications.

\begin{acknowledgments}
B. A. Malomed appreciates hospitality of the State Key Laboratory of
Optoelectronic Materials and Technologies at the Sun Yat-sen
University (Guangzhou, China). This work is supported by the project
of the National Key Basic Research Special Foundation
(G2010CB923204), Chinese National Natural Science Foundation
(10930411, 10774193).
\end{acknowledgments}

\appendix

%

\bibliography{apssamp}

\end{document}